\documentclass{ws-ijmpd}

\begin{document}

\markboth{Gustavo E. Romero, Daniela P\'erez}
{Time and irreversibility in an accelerating universe}

%%%%%%%%%%%%%%%%%%%%% Publisher's Area please ignore %%%%%%%%%%%%%%%
%
\catchline{}{}{}{}{}
%
%%%%%%%%%%%%%%%%%%%%%%%%%%%%%%%%%%%%%%%%%%%%%%%%%%%%%%%%%%%%%%%%%%%%

\title{Time and irreversibility in an accelerating universe\footnote{This article received an Honorable Mention from the Gravity Research Foundation in 2011.}}

\author{Gustavo E. Romero\footnote{Chief Researcher, CONICET, Argentina.}}

\address{Instituto Argentino de Radioastronom{\'{i}}a (IAR, CCT La Plata, CONICET), C.C. No. 5, 1894,\\
Villa Elisa, Buenos Aires, Argentina.\\
FCAyG, Observatorio de La Plata, Paseo del Bosque s/n,
CP 1900 La Plata, Argentina.\\
romero@iar-conicet.gov.ar}

\author{Daniela P\'erez\footnote{Fellow of CONICET, Argentina.}}

\address{Instituto Argentino de Radioastronom{\'{i}}a (IAR, CCT La Plata, CONICET), C.C. No. 5, 1894,\\
Villa Elisa, Buenos Aires, Argentina.\\
danielaperez@iar-conicet.gov.ar}

\maketitle

\begin{history}
\received{Day Month Year}
\revised{Day Month Year}
\comby{Managing Editor}
\end{history}

\begin{abstract}
It is a remarkable fact that all processes occurring in the observable universe are irreversible, whereas the equations through which the fundamental laws of physics are formulated are invariant under time reversal. The emergence of irreversibility from the fundamental laws has been a topic of consideration by physicists, astronomers and philosophers since Boltzmann's formulation of his famous ``H'' theorem. In this paper we shall discuss some aspects of this problem and its connection with the dynamics of space-time, within the framework of modern cosmology. We conclude that the existence of cosmological horizons allows a coupling of the global state of the universe with the local events determined through electromagnetic processes.  \\
\end{abstract}

\section{Introduction}	

\begin{flushright}

\begin{quote}

\vspace{0.5cm}
\hspace{4.5cm}	My days sprint past me like runners,\\
\hspace{4.5cm}	I will never see them again.\footnote{{\sl The Book of Job}, translated by Stephen Mitchell, HarperPrennial, New York, 1992.}\\
\vspace{0.5cm}
\hspace{8.5cm} Job

\end{quote}

\end{flushright}
\vspace{0.7cm}

There is something notorious about the world. It {\em changes}. The past seems to be quite different from the future. We can remember the former and, sometimes, predict the latter. We grow older, not younger. The universe was hotter in the past, and very likely it will become colder in the future. The disorder around us seems to increase. All these facts and many others of the kind are expressed in terms of the Second Law of Thermodynamics: {\em The entropy of a closed system never decreases}. If entropy is denoted by $S$, this law reads:
\begin{equation}
	\frac{dS}{dt}\geq 0.
\end{equation}

In the 1870s, Ludwig Boltzmann argued that the effect of randomly moving gas molecules was to ensure that the entropy of a gas would increase, until it reaches its maximum possible value. This is his famous {\em H-theorem}. Boltzmann was able to show that macroscopic distributions of great inhomogeneity (i.e. of high order or low entropy) are formed from relatively few microstate arrangements of molecules, and were, consequently, relatively improbable. Since physical systems do not tend to go into states that are less probable than the states they are in, it follows that any system would evolve toward the macrostate that is consistent with the largest number of microstates. The number of microstates and the entropy of the system are related by the fundamental formula:
\begin{equation}
	S= k \ln W,
\end{equation}
where $k=10^{-23}$ JK$^{-1}$ is Boltzmann's constant and $W$ is the volume of the phase-space that corresponds to the macrostate of entropy $S$. 

More than twenty years after the publication of Boltzmann's fundamental papers on kinetic theory\cite{bol1}\cdash\cite{bol2}, Burbury\cite{bur1}\cdash\cite{bur2} pointed out that the source of asymmetry in the H-theorem is the assumption that the motions of the gas molecules are independent before they collide and not afterward, if entropy is going to increase. This essentially means that the entropy increase is a consequence of the {\em initial conditions} imposed upon the state of the system. Boltzmann's response was\cite{bol3}: 

\begin{quote}
There must then be in the universe, 
which is in thermal equilibrium as a 
whole and therefore dead, here and 
there, relatively small regions of the 
size of our  world, which during the 
relatively short time of eons deviate 
significantly from thermal equilibrium.  
Among these worlds the state probability 
increases as often as it decreases.        
\end{quote} 

As noted by Price\cite{price}: ``The low-entropy condition of our region seems to be associated entirely with a low-entropy condition in our past.'' This is called the Past Hypothesis.

The probability of the large fluctuations required for the formation of the universe, on the other hand, seems to be zero, as noted long ago by Eddington\cite{ast}: ``A universe containing mathematical physicists 
 at any assigned date will be in the state of 
maximum disorganization which is not inconsistent 
with the existence of such creatures.'' Large fluctuations are rare (the probability $P$ of an entropic variation $\Delta S$ is $P\sim \exp{-\Delta S}$); {\em extremely} large fluctuation, basically impossible. For the whole universe, $\Delta S\sim 10^{104}$ in units of $k=1$ \cite{el}. This yields $P=0$. We are here, however, living momentarily because we are far from thermal equilibrium.   

In this paper we shall discuss a possible source for the existence of local irreversible processes that is related to the presence of cosmological horizons.

\section{Formulation of the problem}

In 1876, a former teacher of Boltzmann and later colleague at the University of Vienna, J. Loschmidt, noted that the laws of (Hamiltonian) mechanics are such that for every solution one can construct another solution by reversing all velocities and replacing $t$ by $-t$\cite{los}. Since the Boltzmann's function $H[f]$ is invariant under velocity reversal, it follows that if $H[f]$ decreases for the first solution, it will increase for the second. Accordingly, the H-theorem cannot be a general theorem for all mechanical evolutions of the gas. More generally, the problem goes far beyond classical mechanics and encompasses our whole representation of the physical world. This is because {\em all formal representations of all fundamental laws of physics are invariant under the operation of time reversal}. Nonetheless, the evolution of all physical processes in the universe is irreversible. 

If we accept, as mentioned in the introduction, that the origin of the irreversibility is not in the laws but in the initial conditions of the equations that represent the laws, two additional problems emerge: 1) what were exactly these initial conditions?, and 2) how the initial conditions, of global nature, can enforce, at any time and any place, the observed local irreversibility? 

The first problem is, in turn, related to the following one, once the cosmological setting is taken into account: in the past, the universe was hotter and at some point matter and radiation were in thermal equilibrium (i.e. in a state of maximum entropy); how is this compatible with the fact that entropy has ever been increasing according to the Past Hypothesis?, how can entropy still increase if it was at a maximum at some past time?  

The standard answer to this question invokes the expansion of the universe: as the universe expanded, the maximum possible entropy increased with the size of the universe, but the actual entropy was left well behind the permitted maximum. The Second Law of Thermodynamics and the source of irreversibility is the trend of the entropy to reach the permitted maximum. According to this view, the universe actually began in a state of maximum entropy, but due to the expansion, it was still possible for the entropy to continue growing\cite{gold}.        

The main problem with this line of thought is that it is not true that the universe was in a state of maximum disorder at some early time. In fact, although locally matter and radiation might have been in thermal equilibrium, this situation occurred in a regime where the  local effects of gravity cannot be ignored.  Penrose\cite{penrose} suggested that entropy might be assigned to the gravitational field itself. Though locally matter and radiation were in thermal equilibrium in the past, the gravitational field should have been quite far from equilibrium, since gravity is an attractive force and the universe was initially structureless. Consequently, the early universe was globally out of equilibrium, being the total entropy dominated by the entropy of the gravitational field. 

In absence of a theory of quantum gravity, a statistical measure of the entropy of the gravitational field is not possible. The study of the gravitational properties of macroscopic systems through classic general invariants, however, might be a suitable approach to the problem. Penrose proposed that the Weyl curvature tensor can be used to specify the gravitational entropy\cite{penrose}. Several prescriptions have been proposed since then to estimate the entropy associated with the classical field, on the basis of scalars constructed out of different functions of the Weyl scalar\cite{scalars}.

\section{Electrodynamics and cosmology}

What makes physical processes occur in a preferred direction of space-time if the physical laws are expressed by time-invariant equations? If entropy globally increases because it was low in the past, how this enforces local changes in a particular sense? 

We suggest that there is a global-to-local relation between the conditions in the far past and future, related to the dynamical state of the universe, with the local physics that determines the way affairs occur in and around us. 

The basic processes in our brain and those we perceive through our senses are of electromagnetic origin. Gravity is far too weak in comparison to electromagnetism. The other fundamental interactions, strong and weak, are of very short range. If gravitational contributions dominate the low entropy in the early universe, there should be some coupling between gravity and electromagnetism that determines the direction along which heat flows.  
   
The electromagnetic radiation field can be described in the terms of a 4-potential $A^{\mu}$, which satisfies linear equations:
\begin{equation}
\partial^{\nu}\partial_{\nu}A^{\mu}(\vec{r},\;t)=4\pi j^{\mu} (\vec{r},\;t), \label{Maxwell}
\end{equation}
where we have considered units such that $c=1$ and $j^{\mu}$ represents the 4-current. The solution $A^{\mu}$ is a functional of the sources $j^{\mu}$. This type of equation admits both retarded and advanced solutions. 
\begin{equation}
	A^{\mu}_{\rm ret}(\vec{r},\;t)=\int_{V_{\rm ret}}
\frac{j^{\mu} \left(\vec{r},\;t-\left|\vec{r}-\vec{r'}\right|\right)}{\left|\vec{r}-\vec{r'}\right|}d^{3}\vec{r'} + \int_{\partial V_{\rm ret}}
\frac{j^{\mu} \left(\vec{r},\;t-\left|\vec{r}-\vec{r'}\right|\right)}{\left|\vec{r}-\vec{r'}\right|}d^{3}\vec{r'}, \label{ret}
\end{equation}
\begin{equation}
	A^{\mu}_{\rm adv}(\vec{r},\;t)=\int_{V_{\rm adv}}
\frac{j^{\mu} \left(\vec{r},\;t+\left|\vec{r}-\vec{r'}\right|\right)}{\left|\vec{r}-\vec{r'}\right|}d^{3}\vec{r'} + \int_{\partial V_{\rm adv}}
\frac{j^{\mu} \left(\vec{r},\;t+\left|\vec{r}-\vec{r'}\right|\right)}{\left|\vec{r}-\vec{r'}\right|}d^{3}\vec{r'}. \label{adv}
\end{equation}

The two functionals of $j^{\mu}(\vec{r}, t)$ are related to one another by a time
reversal transformation. The solution (\ref{ret}) is contributed by sources in the
past of the space-time point $p(\vec{r}, t)$ and the solution (\ref{adv}) by sources in the
future of that point. The integrals in the second term on the right side are
the surface integrals that give the contributions from i) sources outside of $V$
and ii) source-free radiation. If $V$ is the causal past ($J^{-}$) and future ($J^{+}$), the surface
integrals do not contribute since material sources both outside $V$ and on the boundary are causally disconnected from $p(\vec{r}, t)$. We also assume Sommerfeld radiation condition, that makes source-free radiation null. 

The linear combinations of electromagnetic solutions are also solutions,
since the equations are linear and the Principle of Superposition holds. It is
usual to consider only the retarded potential as physical meaningful in order
to estimate the electromagnetic field at $p(\vec{r}, t)$: $F^{\mu\nu}_{\rm ret}=\partial^{\mu} A^{\nu}_{\rm ret}- \partial^{\nu}A^{\mu}_{\rm ret}.$ There seems to be no compelling reason, however, for such a choice\footnote{ See, for instance, references \refcite{Fokker}, \refcite{Dirac}, \refcite{W-F1}, \refcite{W-F2}, and \refcite{clarke}.}. We can
adopt, for instance (in what follows we use a simplified notation),

\begin{equation}
	A^{\mu}(\vec{r}, t)=\frac{1}{2}\left(\int_{J^{-}} {\rm ret} +\int_{J^{+}} {\rm adv}\right) dV.
\end{equation}

If the sources in the past and future are the same, and the boundary conditions are the same, both solutions are identical. Given the dynamical state of the universe, characterized by an accelerated expansion, the causal past and future of a point $p(\vec{r},\;t)$ are not, however, necessary symmetric in what the number of charges contained concerns.
   
If the space-time is curved, the null cones that determine the local causal structure will not be symmetric around the point $p(\vec{r},\;t)$. In particular, the presence of cosmological particle horizons can make very different the contributions of both solutions. Particle horizons occur whenever a particular system never gets to be influenced by the whole space-time. If a particle crosses the horizon, it will not exert any further action upon the system respect to which the horizon is defined. 

Finding the particle horizons (if one exists at all) requires a knowledge of the global space-time geometry. Particle horizons occur in systems undergoing lasting acceleration. 

The radius of the past particle horizon is\cite{rindler}:

\begin{equation}
	R_{\rm past}= a(t) \int^{t}_{t'=0} \frac{c}{a(t')} dt',
\end{equation}
where $a(t)$ is the time-dependent scale factor of the universe. The radius of the future particle horizon (sometimes called event horizon) is:

\begin{equation}
	R_{\rm future}= a(t_0) \int^{\infty}_{t'_0} \frac{c}{a(t')} dt'.
\end{equation}

If the universe is accelerating, as it seems to be suggested by recent observations\cite{permu}, then $J^{+}(p)$ and $J^{-}(p)$ are not symmetric because of the presence of future horizons. This implies that $A^{\mu}_{\rm ret}$ and $A^{\mu}_{\rm adv}$ will be different.   
We can then introduce a vector field $L^{\mu}$ given by:  
   
\begin{equation}
L^{\mu}= \left[ \int_{J^{-}}{\rm ret}- \int_{J^{+}}{\rm adv}\right] dV\neq 0.
\end{equation}

If $g_{\mu\nu}L^{\mu}T^{\nu}\neq0$, with $T^{\nu}=(1,0,0,0)$, there is a preferred direction for the flux of electromagnetic energy in space-time. If the sign of $T$ is chosen in such a way that it is positive in the direction of the global expansion of the universe, the electromagnetic flux will go from what we call past to future if $L>0$, i.e. if there is a future particle horizon hidden some electromagnetic currents. The (Poynting) flux is given by:

\begin{equation}
	\vec{S^{\mu}}=( \vec{E}^{2}+\vec{B}^{2}, \vec{E} \times \vec{B})=(T^{00}_{\rm EM}, \;T^{01}_{\rm EM},\; T^{02}_{\rm EM},\; T^{03}_{\rm EM}),
\end{equation}
where $\vec{E}$ and $\vec{B}$ are the electric and magnetic fields, both determined from $A^{\mu}$, and $T^{\mu\nu}_{\rm EM}$ is the electromagnetic energy-momentum tensor.

In a black hole interior the direction of the Poynting flux is toward the singularity at its center. In an expanding, accelerating universe, it is in the global future direction.  Then, the fact that there is a time-like vector field along where Poynting flux occurs, indicates the existence of a future particle horizon. There is a global-to-local relation given by the Poynting flux as determined by the curvature of space-time that indicates the direction along which events occur. Physical processes, inside a black hole, occur along a different direction from outside. The causal structure of the world is determined by the dynamics of space-time and the initial conditions. Macroscopic irreversibility\footnote{The electromagnetic flux is related with the macroscopic concept of temperature through the Stefan-Boltzmann law: $L=A\sigma_{\rm SB}T^{4}$, where $\sigma_{\rm SB}= 5.670 400 \times 10^{-8} \textrm{J\,s}^{-1}\textrm{m}^{-2}\textrm{K}^{-4}$ is the 
Stefan-Boltzmann constant.} emerges from fundamental reversible laws. 

There is an important corollary to these conclusions. Local observations about the direction of events can provide information about global features of space-time and the existence of horizons and singularities. 

\section{Causal explanations?}

Do the initial conditions of the universe, namely the fact that the gravitational entropy was extremely small, require a causal explanation? What such an explanation would be?  

The causal relation is a relation between events (ordered pairs of states), not between things. Causation is a form of event generation\cite{be,bu}. The initial conditions represent a state of a thing (the universe in this case, the maximal thing) and hence have no causal power. The initial conditions are a ``state of affairs''. The causal power should be looked for in previous events, but if space-time itself, as an emergent property of basic things, has a quantum behavior, classical causality would not operate. Rather, the initial conditions should appear as a classical limit of the gravitational processes at quantum level. Final conditions can be causally explained because there are possible causes that precede final conditions, but initial conditions, on the contrary, cannot be causally explained because there is no time that precede them. The initial conditions of the universe, then, should have an explanation in terms of yet unknown dynamical laws. Such laws do not need, and likely have not, a causal structure.

\section{Final remarks}

Time is an emergent property of changing things. It is represented by a one-dimensional continuum. Processes in space-time are anisotropic, although physical laws are invariant under time reversal. Time itself is not anisotropic, because it is not represented by a vector field. For instance, in a Friedman-Robertson-Walker-Lema\^{\i}tre model, time is represented by the following real parameter\cite{rindler}:
$$
t=\int\frac{da}{\sqrt{F(a)}},
$$
where $a$ is the cosmic scale factor, $F(a)={C}/{a}+(\Lambda a^{2})/3 - k$, $C=(8/3)\pi G\rho a^{3}$, and the remaining symbols have their usual meaning in the literature\cite{rindler}. This time coincides with proper time along each fundamental worldline, clearly being described by a real number.
 
Our conclusion is that the dynamical state of space-time and the initial conditions determine the local direction of the physical processes through the electromagnetic Poynting flux. There is a global-to-local relation between gravitation and electrodynamics, between the origin and fate of the universe, and processes such as those in our brains.  From the irreversibility observed around us we can infer the existence of a future cosmological particle horizon. \\

\section*{Acknowledgments}

This work has been supported by Grant CONICET PIP 0078. GER thanks Mario Bunge for stimulating comments.


\begin{thebibliography}{0}    %for 1 digit

\bibitem{bol1}L. Boltzmann, {\it Wiener Berichte} {\bf 66} (1872) 275.
\bibitem{bol2}L. Boltzmann, {\it Wiener Berichte} {\bf 76} (1877) 373.
\bibitem{bur1}S.H. Burbury, {\it Nature} {\bf 51} (1894) 78.
\bibitem{bur2}S.H. Burbury, {\it Nature} {\bf 51} (1895) 320.
\bibitem{bol3}L. Boltzmann, {\it Nature} {\bf 51} (1895) 413.   
\bibitem{price} H. Price, {\it Contemporary Debates in Philosophy of Science}, ed.~C. Hitchcock (Blackwell, Singapore, 2004), p.~219.   
\bibitem{ast}A.S. Eddington, {\it Nature} {\bf 127}, 3203 (1931) 447.       
\bibitem{el} C.A. Egan and H. Lineweaver, {\it The Astrophysical Journal} {\bf 710} (2010) 1825. 
\bibitem{los}J. Loschmidt, {\it Wiener Berichte} {\bf 73} (1876) 128. 
\bibitem{gold}T. Gold, {\it American Journal of Physics} {\bf 30}, 6 (1962) 403. 
\bibitem{penrose}R. Penrose, in {\it General Relativity: An Einstein Centennial}, eds.~S.W. Hawking and W. Israel (Cambridge University Press, Cambridge, 1979), p.~581. 
\bibitem{scalars} $\O$ Rudjord,  $\O $ Gr$\o$n, H. Sigbj$\o$rn, H., {\it Phys. Scr.} {\bf 77} {\it Issue 5}, (2008) 055901, 1-7
%\bibitem{sciama}D. Sciama, in {\it The Nature of Time}, eds.~T. Gold and D.L. Schumacher (Cornell University Press, Ithaca, 1967), pp.~55-67. 
\bibitem{Fokker} A.D. Fokker, {\it Z. Phys.} {\bf 58} (1929) 386.
\bibitem{Dirac}P.A.M. Dirac, {\it Proc. Royal Soc. Lond.} {\bf 167} (1938) 148
\bibitem{W-F1} J.A. Wheeler and R.P. Feynman, {\it Rev. Mod. Phys.} {\bf 17} (1945) 157. 
\bibitem{W-F2} J.A. Wheeler and R.P. Feynman, {\it Rev. Mod. Phys.} {\bf 21} (1949) 425. 
\bibitem{clarke}C.J.S. Clarke, Foundations of Space-Time Theories, in {\it Minnesota Studies in the Philosophy of Science}, Vol.~8 (University of Minnesota Press, Minneapolis, 1977), pp.~94-108. 
\bibitem{rindler}W. Rindler, {\it Monthly Notices of the Royal Astronomical Society} {\bf 116} (1956) 662.
\bibitem{permu}S. Perlmutter et al., {\it The Astrophysical Journal} {\bf 517} (1999) 565.
\bibitem{be}D. P\'erez and G.E. Romero, {\it Bolet{\'\i}n de la Asociaci\'on Argentina de Astronom{\'\i}a}  {\bf 52} (2009) 221.
\bibitem{bu} M. Bunge, {\it Causality in Modern Science}, (Dover, NY, 1979).
\end{thebibliography}
\end{document}